\documentclass[%
 reprint,
 amsmath,amssymb,
 aps,
]{revtex4-1}

\usepackage{graphicx}
\usepackage{dcolumn}
\usepackage{bm}

\usepackage{xr-hyper}
\usepackage{hyperref}

\begin{document}

\preprint{APS/123-QED}

\title{Measuring Topological Constraint Relaxation in Ring-Linear Polymer Blends}

\author{Daniel L. Vigil,$^1$ Ting Ge,$^2$ Michael Rubinstein,$^3$ Thomas C. O'Connor,$^4$ and Gary S. Grest$^1$}
\affiliation{$^1$Sandia National Laboratories, Albuquerque, NM 87185, USA}
\affiliation{$^2$Department of Chemistry and Biochemistry, University of South Carolina, Columbia, South Carolina 29208, USA
}
\affiliation{
$^3$Department of Mechanical Engineering and Materials Science, Duke University, Durham, North Carolina 27708, USA}
\affiliation{$^4$Department of Materials Science and Engineering, Carnegie Mellon University, Pittsburgh, Pennsylvania 15213}%
\date{\today}

\begin{abstract}

Polymers are an effective test-bed for studying topological constraints in condensed matter due to a wide array of synthetically-available chain topologies. When linear and ring
polymers are blended together, emergent rheological properties are observed as the blend can be more
viscous than either of the individual components. This emergent behavior arises since
ring-linear blends can form long-lived topological constraints as the linear polymers thread the ring polymers.
Here, we demonstrate how the Gauss linking integral can be used to efficiently evaluate the relaxation of topological constraints in ring-linear polymer blends.
For majority-linear blends, the relaxation rate of topological constraints depends primarily on reptation of the linear polymers, resulting in the diffusive time $\tau_{d,R}$ for rings of length $N_R$ blended with linear chains of length $N_l$ to scale as
$\tau_{d,R}\sim N_R^2N_L^{3.4}$.

\end{abstract}

\maketitle

Topological constraints are long-lived interactions between atomic degrees of freedom that arise from the entanglement of some element of their phase spaces. They drive a variety of exotic nonlinear dynamics across an array of fields: stabilizing solitons in nonlinear optics \cite{Leach2004,Irvine2010}, hydrodynamic vortices in both classical \cite{Thomson1868,Thomson1869,Kleckner2013} and quantum fluids \cite{Proment2012}, and fractional electronic states in topological insulators and quantum spin liquids.
For most of these systems, the origin of the topological entanglement is subtle and difficult to visualize or characterize directly. 
The dynamics of polymer melts and blends are also dominated by topological entanglement, but unlike many quantum systems, this entanglement arises from the interweaving, threading, and knotting of the polymer chains in real space.
Additionally, the mathematical structure of models of polymer melts is nearly identical to that for many quantum systems \cite{Chandler1981,Fredrickson2023}, so much so that some quantum problems are simulated using ring polymers \cite{Craig2004}. In addition, polymers have the advantage of a large body of synthetic and characterization data compared to many other fields due to advances that allow for synthesis of polymers with nearly arbitrary size and chain topology.
This makes entangled polymer melts an ideal and economical test bed for exploring the dynamics of systems with complex topological constraints.

\begin{figure}[h]
    \centering
    \includegraphics[width=0.9\columnwidth]{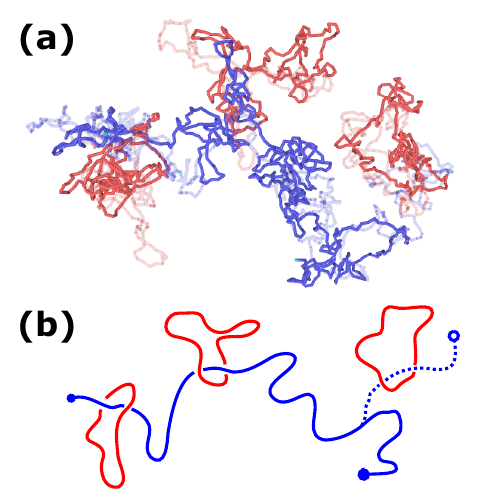}
    \caption{\label{fig:snapshot} (a) Snapshot of three ring polymers (red, $N_R=200$) threaded by a linear polymer (blue, $N_L=600$) from a molecular dynamics simulation. The initial configuration is faded and the final configuration is shown in bold. In the final configuration, the right ring is no longer threaded. (b) Drawing of the dethreading of the right ring polymer due to motion of the linear chain. The initial configuration is shown with a dashed line and the final configuration is shown with the solid line. Chain ends are marked with circles.}
\end{figure}

For melts of linear polymers, there are mature techniques to evaluate topological constraints. These include contour reduction algorithms \cite{Sukumaran2005,Tzoumanekas2006} and isoconfigurational averaging \cite{Bisbee2011}. Topological constraints in ring polymers, in contrast, have been more difficult to measure. Additionally, recent studies have found that neat ring polymers have significantly different rheological properties from linear polymers due to the differences in knotting and entanglement \cite{Koniaris1991,halverson11b,Doi2015,Ge2016,Parisi2021a,ubertini2022,Tu2023,stavno2023}. When ring and linear polymers are blended together, emergent rheological properties are observed as the blend can be more viscous than either of the individual components \cite{Roovers1988,Peddireddy2020,Parisi2021}. This emergent behavior has been ascribed to the fact that ring-linear blends can form topological constraints via linear polymers threading the ring polymer (Fig. \ref{fig:snapshot}) and these ring-linear threads are presumed to be long-lived. Thus far, direct observation of the ring-linear threading/dethreading process has been difficult in experiments and simulations. In this work we use recently implemented topology tools to directly measure ring-linear thread relaxation in simulations of ring-linear polymer blends.

We perform coarse-grained molecular dynamics (MD) simulations of polymer melts where individual polymers are modeled by bead-spring chains with FENE bonds and all beads interact via purely repulsive Lennard-Jones interactions characterized by energy $\epsilon$ and distance $\sigma$. Model details are presented in the SI. Linear chains contain $N_L$ beads and rings contain $N_R$ beads. All simulations are conducted in cubic cells with periodic boundary conditions at a particle density of 0.85 $m/\sigma^3$, where $m$ is the mass of a bead. Unconcatenated ring polymers were constructed according to previously published methods \cite{Smrek2019} and blends of various ring volume fraction $\phi_R$ were constructed by removing a bond from some rings to convert them into linear chains. Simulations are conducted with a Langevin thermostat at temperature $T=\epsilon/k_B$ with damping parameter 100$\tau$ and were time integrated using a velocity-Verlet algorithm with time step $0.01\tau$, where $\tau=\sqrt{m\sigma^2/\epsilon}$ is the Lennard-Jones time. All simulations were conducted using LAMMPS \cite{Thompson2022}. System sizes and equilibration times are given in the SI; the blends contained up to 960000 particles and were simulated for up to 10 billion time steps ($10^8 \tau$).

To evaluate ring-linear threads, we use an approach based on the Gauss linking integral (GLI), which has recently been implemented in a parallel, open-source code TEPPP \cite{Herschberg2023}. As a post-processing step of our simulations, the periodic linking number $L_P$ (a generalization of the linking number to periodic simulations \cite{Panagiotou2011,Panagiotou2015}) is computed between all pairs of ring and linear chains. Any pair of chains with $|L_P|>0.5$ is consider threaded. The periodic linking number can take any real value as we do not invoke a closure approximation on the linear chain, unlike previous work \cite{Hagita2021}. All beads from the linear chain are included in the analysis, unlike previous work which excluded beads that were within one entanglement length $N_e$ of the chain end \cite{OConnor2022}. To further characterize the time dependence of threads, we construct a thread correlation function, $C(t)$, analogous to the intermittent association correlation function used to study ion associations in solution \cite{Luzar1996,Muller1998,Chandra2000,Zhao2009}. Details of the linking number calculation, thread cutoff, and correlation function are discussed further in the SI. The number of ring-linear threads has also been counted via other techniques such as primitive path analysis with contact mapping \cite{OConnor2022}, minimal surfaces \cite{Smrek2016,Smrek2019,Wang2024}, and persistent homology \cite{Landuzzi2020}, though most of these authors have not been able to measure the dynamics of threading nor multiple threads.

\begin{figure}
    \centering
    \includegraphics{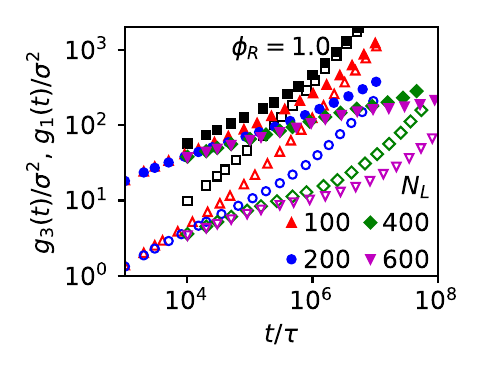}
    \caption{Mean-squared displacement of ring polymers of length $N_R=400$ in pure melts (black squares) or blends with $\phi_R=0.3$ (all other symbols). Mean squared displacement of the center of  mass $g_3(t)$ is shown with open symbols and motion of a monomer $g_1(t)$ with filled symbols.}
    \label{fig:ringMSD}
\end{figure}

The dynamics of individual ring and linear polymers are characterized by the diffusion time $\tau_d$. Here we define $\tau_d$ as the time for the mean squared displacement (MSD) of a bead  $g_1(\tau_d)=\langle (\Delta r(\tau_d))^2 \rangle$ to move $3 \langle R_g^2 \rangle$, where $R_g$ is the radius of gyration of a chain \cite{Hsu2016}. An example of the MSD of the center of mass $g_3(t)$ motion and of a bead $g_1(t)$ are presented in Fig.~\ref{fig:ringMSD} for rings of length $N_R=400$ in a pure ring melt and in a blend with $\phi_R=0.3$ with varying length of the linear chains $100\le N_L\le 600$. Fig.~\ref{fig:ringMSD} clearly shows how as $N_L$ increases, the ring motion becomes subdiffusive as the rings have to `wait' for the linear chains to release the topological constraints before the ring can relax. While the motion of the rings depends strongly on the length of the linear chains, the motion of the linear is hardly affected by the presence of the rings as shown in Fig.~S6.

Theories for chain dynamics in ring-linear blends posit that rings that are threaded by linears cannot diffuse until the topological constraint imposed by the thread is released via linear chain reptation (as seen in the MSD plots). The classical constraint release model assumes that there are $N_R/N_e$ threads per ring and that the threads are released independently \cite{Parisi2020}. The time scale for an individual relaxation event scales with the linear chain diffusion time, which scales with the linear chain size as $\tau_{d,L}\sim N_L^{3.4}$. The Rouse-like constraint release time for the ring polymer then scales like $\tau_{d,R}\sim N_R^2N_L^{3.4}$. If the linear chains are short, however, there will be a crossover to unentangled ring Rouse relaxation where the diffusion time of the ring scales like $\tau_{d,R} \sim N_R^2$ and is independent of linear chain size.

\begin{figure}[h]
\includegraphics{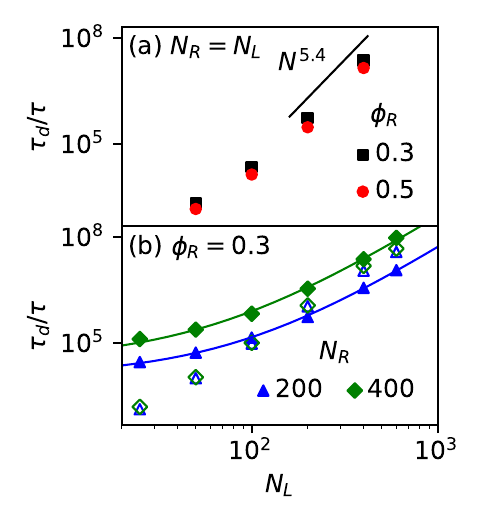}
\caption{\label{fig:diffusion} Diffusion times for ring (filled symbols) and linear polymers (open symbols) versus linear chain length $N_L$ in a ring-linear blend with (a) equal ring and linear polymer length, (b) fixed ring polymer length and ring fraction $\phi_R=0.3$. Solid lines indicate fitted crossover functions from eq S9.
}
\end{figure}

\begin{figure*}
\includegraphics{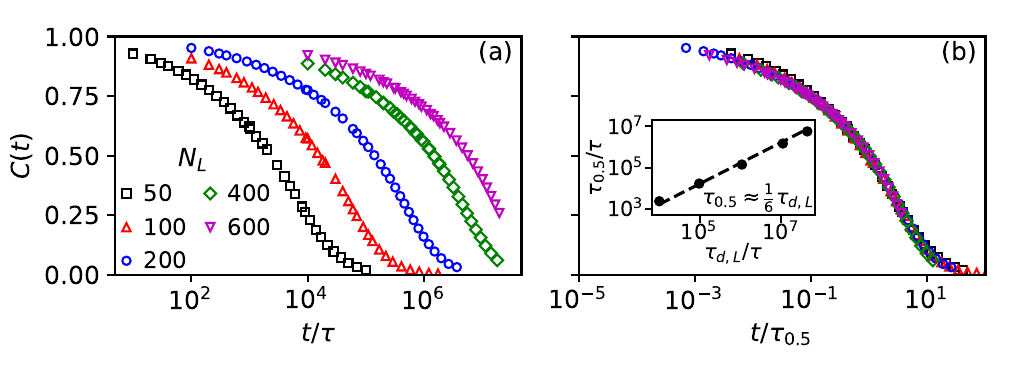}
\caption{\label{fig:linthread} (a) Dethreading correlation function $C(t)$ for a ring-linear blend with ring fraction $\phi_R=0.3$ and ring length $N_R=200$. (b) $C(t)$ versus $t/\tau_{0.5}$, where $\tau_{0.5}$ is the time at which only 50\% of the original ring-linear threads remain, which is different for each linear length. The inset shows the thread relaxation time scale $\tau_{0.5}$ versus the diffusion time for a linear chain, $\tau_{d,L}$. }
\end{figure*}

\begin{figure}
\includegraphics{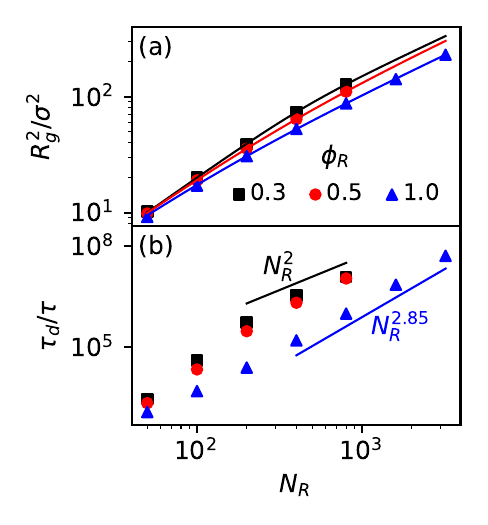}
\caption{\label{fig:ringdiff} (a) Radius of gyration $R_g$ and (b) diffusion time $\tau_d$ for ring polymers versus ring chain length $N_R$ in a ring-linear blends and a pure ring melt. In the blends, the linear chain length $N_L=200$. Solid lines in part (a) indicate fit to crossover functions given in eq S10.
}
\end{figure}

The diffusion time for ring polymers in blends with linear polymers of equal chain length $N_R=N_L=N$ as a function of linear chain length is shown in Fig.~\ref{fig:diffusion}a. The diffusion time $\tau_d$ of ring polymers increases as $N^{5.4}$, in agreement with theory. In contrast, pure ring polymers have a diffusion time that can be fit to an apparent power law $\tau_d\sim N_R^{2.8}$ (Fig. \ref{fig:ringdiff}b) and pure, entangled linear polymers have a diffusion time that scales as $N_L^{3.4}$\cite{halverson11b}.

We next examine the effect of the linear chain size on the relaxation of the rings. We fix the ring size ($N_R=200$ or $N_R=400$) and vary the linear chain length as shown in Fig.~\ref{fig:diffusion}b. The open symbols show the diffusion time of the linear chains, which scales like $N_L^{3.4}$, which is expected for entangled linear polymers. For the model considered here, the linear chain entanglement length is $N_e\approx28$ \cite{Hsu2016}. For sufficiently long linear polymers, the ring polymer diffusion time also follows the $N_L^{3.4}$ scaling. However, for short linear chains, the ring polymer diffusion has a weaker dependence on $N_L$ and can be fit to a crossover to unentangled ring polymer scaling $\tau_{d,R}\sim N_L^0$. The solid lines are a fit to a crossover function (eq S9
) that includes the $N_L^{3.4}$ and $N_L^0$ limits. The fit indicates a crossover $N_L$ value around 80.

 The thread relaxation $C(t)$ for blends with $\phi_R=0.3$, $N_R=200$, and variable $N_L$ is shown in Fig.~\ref{fig:linthread}a. $C(t)$ shows a similar shape for all $N_L$, but with shifted time scales. 
 These data can be collapsed by choosing a value of the correlation function, in this case 0.5, and rescaling time for each curve so that all curves overlap at the chosen value. The time-rescaled data is given in Fig.~\ref{fig:linthread}b and the inset shows the times $\tau_{0.5}$ used to collapse all the data versus the diffusion time $\tau_{d,L}$ of the linear chains. The $C(t)$ curves collapse nearly perfectly onto each other, indicating that the dethreading dynamics is similar with increasing linear polymer size. 
 
 The inset shows that $\tau_{0.5}$ is directly proportional to the linear chain diffusion time, though it is smaller by a factor of $\approx1/6$ for blends with $\phi_R=0.3$ and $N_R=200$. To relax the thread only requires a portion of the linear chain to reptate through the ring, so it is expected for the dethreading time to be less than the diffusion time.

We now evaluate the effect of the ring polymer size on the chain diffusion and dethreading.
When varying $N_R$ one must be careful to note that small rings and large rings in pure ring melts have different scaling behavior. In pure ring melts, small rings are almost unperturbed Gaussian rings with size that scales like $R_g^2 \sim N$. As the size of the ring increases, the rings impinge on each other and there is a crossover to a loopy globule scaling regime where $R_g^2 \sim N^{2/3}$ \cite{Kruteva2023}. 
 Rings in ring/linear blends are expected to follow $R_g^2 \sim N_R$ if the rings are sufficiently diluted by linears. If a critical ring concentration is exceeded in the blend, then the large ring scaling $R_g^2 \sim N_R^{2/3}$ will be recovered. The critical concentration is a function of the ring size, so increasing ring polymer size at fixed volume fraction of rings may cause one to cross the critical concentration.

Fig.~\ref{fig:ringdiff}a shows the mean squared radius of gyration of ring polymers in ring-linear blends ($\phi_R=0.3$ and $\phi_R=0.5$) and pure ring melts ($\phi_R=1.0$) versus the size of the rings $N_R$. The linear chains have length $N_L=200$ in the blends.
 The data in Fig.~\ref{fig:ringdiff}a were fit to a crossover function (eq S10)
and a crossover $N_R$ was extracted for the blends and the pure rings. For blend systems the crossover occurs for $N_R=445$ ($\phi_R=0.3$) or $N_R=328$ ($\phi_R=0.5$), so most of the data lies in the dilute-ring scaling regime. For the pure ring melts ($\phi_R=1$) the crossover occurs for $N_R=122$, so most of the data lies in the concentrated ring regime.

\begin{figure}
\includegraphics{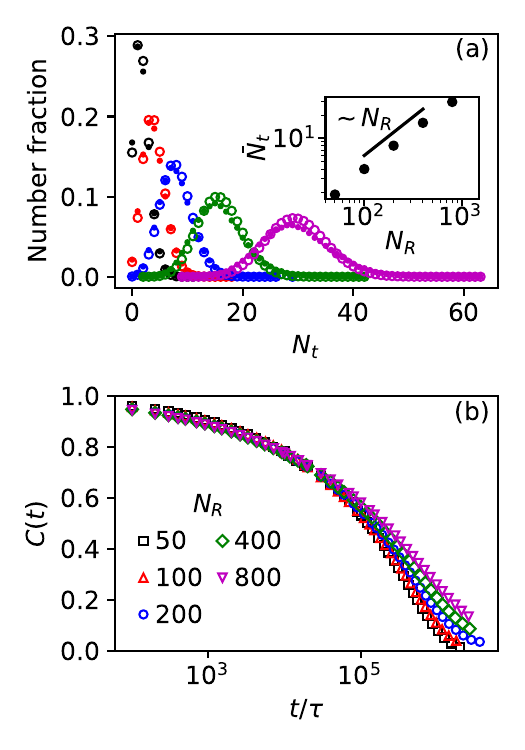}
\caption{\label{fig:ringthread} (a) Number fraction of ring polymers with a given number of linear polymers threading the ring, $N_t$. Linear polymers have length $N_L$ = 200 and  $\phi_R=0.3$. Solid dots indicate results from MD simulations. Open symbols indicate a Poisson distribution with identical mean to the MD results. Inset shows the average number of linear chains threading the ring $\bar{N}_t$ versus ring polymer size $N_R$. (b) Dethreading correlation function $C(t)$ versus time $t$ for the same blends as in (a).}
\end{figure}

The diffusion times for ring polymers are shown in Fig.~\ref{fig:ringdiff}b. For rings that are of similar size to the linear chains ($100\leq N_R \leq 400$), the ring diffusion time in the blends (red and black points) is an order of magnitude larger than in the pure melt ring (blue points). In this regime the ring motion is dominated by ring-linear threadings, which are slow to relax. For small rings ($N_R\leq50$) the diffusion times in blends  is much closer to the diffusion time in the pure ring melt. This is because the rings are so small that they have zero to two threads (Fig. \ref{fig:ringthread}a), and thus have few topological constraints to slow them down compared to the pure ring melt. 
For larger rings ($N_R\geq 800$), the blend and pure melt diffusion times also approach in value. In this regime, the ring dynamics are also affected by ring-ring interactions which are similar between blends and pure ring melts.

The number fraction of rings with a given number of linear chains threading the ring, $N_t$, is shown in Fig.~\ref{fig:ringthread}a. The solid points show the value measured in MD simulations via the GLI analysis. Open symbols indicate a Poisson distribution with mean equal to that of the MD results. For the smallest ring size $N_R=50$, the majority of rings have two or fewer linear chains threading, and 17\% of rings have no linear chains threading them at all. As the ring size $N_R$ is increased, the distribution of number of threads broadens and moves to larger mean values, consistent with previous investigations \cite{Hagita2021,Wang2024}. Notably the Poisson distribution fits the measured distribution well for all $N_R$. This may indicate that ring-linear threads are independent of each other. The inset of Fig.~\ref{fig:ringthread}a shows the average number of linear chains threading a ring, $\bar{N}_t$, versus the ring size $N_R$. The average number of linears threading a ring increases linearly with the ring size, $\bar{N}_t \sim N_R$, which is consistent with previous results based on primitive path analysis \cite{OConnor2022} and minimal surfaces \cite{Wang2024}.

 The dethreading correlation function for the blends with $\phi_R = 0.3$ is shown in Fig.~\ref{fig:ringthread}b. For times $t/\tau < 10^5$ the curves overlap. Note that no rescaling of time has been performed, unlike in Fig.~\ref{fig:linthread}b. The overlapped curves indicate that dethreading dynamics at short times is largely independent of ring size. This indicates that it is the motion of the linear that largely drives dethreading, consistent with previous work.

At long times, $C(t)$ decays to zero more quickly for smaller rings whereas larger rings have a slow relaxing component that gets slower with increasing ring size. The universal functional form that was observed for linear polymers of different sizes in the blends does not occur for rings.
Thus, blending ring and linear polymers has an asymmetric effect where thread relaxation has some complicated dependence on ring polymer size and diffusion, but for linear chains depends only on the diffusion time. 
\smallskip


\begin{acknowledgments}
D.L.V. acknowledges useful discussion with Eleni Panagiotou, the author of the TEPPP software. T.G. acknowledges start-up funds from the University of South Carolina. T.O. acknowledges start-up funds from Carnegie Mellon University.
M. R. acknowledges financial support from the National Institutes of Health under Grant P01-HL164320.
This work was supported in part by the National Science Foundation EPSCoR Program under NSF Grant Np.~OIA-1655740.
Any opinions, findings, and conclusions, or recommendations expressed in this material are those of the authors and do not necessarily reflect those of the National
Science Foundation. 
This work was performed, in part, at the Center for Integrated Nanotechnologies, an Office of Science
User Facility operated for the U.S. Department of Energy (DOE) Office of Science. Sandia National Laboratories is a multi-mission laboratory managed and operated by
National Technology \& Engineering Solutions of Sandia, LLC, a wholly owned subsidiary of Honeywell International, Inc., for the U.S. DOE’s National Nuclear Security Administration under Contract No.DE-NA-0003525.
The views expressed in the Letter do not necessarily represent the views of the U.S. DOE or the U.S. Government.

\end{acknowledgments}

%

\end{document}


\title{Supplemental Information: \break
Measuring Topological Constraint Relaxation in Ring-Linear Polymer Blends}

\author{Daniel L. Vigil,$^1$ Ting Ge,$^2$ Michael Rubinstein,$^3$ Thomas C. O'Connor,$^4$ and Gary S. Grest$^1$}

\affiliation{$^1$Sandia National Laboratories, Albuquerque, NM 87185, USA}
\affiliation{$^2$Department of Chemistry and Biochemistry, University of South Carolina, Columbia, South Carolina 29208, USA
}
\affiliation{
$^3$Department of Mechanical Engineering and Materials Science, Duke University, Durham, North Carolina 27708, USA}
\affiliation{$^4$Department of Materials Science and Engineering, Carnegie Mellon University, Pittsburgh, Pennsylvania 15213
}%
\maketitle

\section{Polymer model}
Polymers are composed of beads of mass $m$ that interact through Lennard-Jones interactions
\begin{equation}
    V_\mathrm{LJ}(r) = 4\epsilon\left[\left(\frac{\sigma}{r}\right)^{12} -\left(\frac{\sigma}{r}\right)^{6}  \right]
\end{equation}
The interaction applies to all non-bonded beads with separation $r<r_c=2^{1/6}\sigma$ so that beads are purely repulsive. Polymers are connected by FENE bonds \cite{Kremer1990} whose interaction is 
\begin{equation}
    V_\mathrm{FENE}(r) = V_\mathrm{LJ}(r)-\frac{KR_0^2}{2}\ln\left[1-\left(\frac{r}{R_0}\right)^2\right]
\end{equation}
In this work we use $K=30\epsilon$ and $R_0=1.5\sigma$. The polymers are made semiflexible by a cosine angle interactions with potential
\begin{equation}
    V_\mathrm{angle}(\theta) = k_\theta \left[1+\cos(\theta)\right]
\end{equation}
We use $k_\theta = 1.5\epsilon$ for all calculations in this work.
\newpage
\section{System sizes}
All of the ring, linear blends simulated for this study are listed in Table \ref{tab:sizes}. The details for the pure ring simulations ($\phi_R=1$) can be found in the supporting information of Tu  et al.~\cite{Tu2023}.

 \begin{table}[h]
      \caption{Ring-linear blends simulated for this study, including ring volume fraction ($\phi_R$), chain lengths ($N_R$ and $N_L$), number of chains ($M_R$ and $M_L$) and total run time $\tau_\mathrm{run}$}
     \label{tab:sizes}
     \centering
     \begin{tabular}{c|S[table-format=4.0]|S[table-format=3.0]|S[table-format=4.0]|S[table-format=5.0]|S[table-format=3.1]}
        \hline
        $\phi_R$ & {$N_R$} & {$N_L$} & {$M_R$} & {$M_L$} & {$t_\mathrm{run}/(10^6\tau)$} \\
        \hline
         0.3 & 50 & 50 & 2400 & 5600 & 1 \\
         0.3 & 50 & 200 & 2400 & 1400 & 5.7 \\
         0.3 & 100 & 100 & 1200 & 2800 & 2.7 \\
         0.3 & 100 & 200 & 1200 & 1400 & 2.4 \\
         0.3 & 200 & 25 & 600 & 11200 & 0.2 \\
         0.3 & 200 & 50 & 600 & 5600 & 0.5 \\
         0.3 & 200 & 100 & 600 & 2800 & 4.2 \\
         0.3 & 200 & 200 & 600 & 1400 & 39.2 \\
         0.3 & 200 & 400 & 600 & 700 & 36 \\
         0.3 & 200 & 600 & 600 & 466 & 63.2\\
         0.3 & 400 & 25 & 384 & 14464 & 4 \\
         0.3 & 400 & 50 & 384 & 7232 & 9 \\
         0.3 & 400 & 100 & 384 & 3616 & 18\\
         0.3 & 400 & 200 & 384 & 1808 & 31 \\
         0.3 & 400 & 400 & 384 & 904 & 86 \\
         0.3 & 400 & 600 & 384 & 602 & 116 \\
         0.3 & 800 & 100 & 360 & 6720 & 10.2 \\
         0.3 & 800 & 200 & 360 & 3360 & 32.5 \\
         0.3 & 1600 & 100 & 180 & 6720 & 24 \\
         0.5 & 50 & 50 & 4000 & 4000 & 10 \\
         0.5 & 50 & 200 & 4000 & 1000 & 7.4 \\
         0.5 & 100 & 100 & 800 & 800 & 8 \\
         0.5 & 100 & 200 & 800 & 400 & 13 \\
         0.5 & 200 & 25 & 1000 & 8000 & 3 \\
         0.5 & 200 & 50 & 1000 & 4000 & 7.9\\
         0.5 & 200 & 100 & 1000 & 2000 & 7.5\\
         0.5 & 200 & 200 & 800 & 800 & 18 \\
         0.5 & 200 & 400 & 800 & 400 & 20.4\\
         0.5 & 200 & 600 & 800 & 266 & 41 \\
         0.5 & 400 & 25 & 800 & 12800 & 2 \\
         0.5 & 400 & 50 & 904 & 7232 &  7\\
         0.5 & 400 & 100 & 904 & 3616 & 9 \\
         0.5 & 400 & 200 & 904 & 1808 & 22 \\
         0.5 & 400 & 400 & 904 & 904 & 81 \\
         0.5 & 800 & 100 & 600 & 4800 & 33\\
         0.5 & 800 & 200 & 600 & 2400 & 45 \\
         0.5 & 800 & 400 & 600 & 1200 & 76 \\
         0.5 & 1600 & 100 & 300 & 4800 & 20\\
        \hline
     \end{tabular}

 \end{table}

\newpage
\section{Linking number}


The linking number $L$ is a measure of how entwined two curves are in space. If both curves are closed then the linking number can be computed by counting the number of times the curves cross from any projection and then adding the number of ``positive'' crossing and subtracting the number of ``negative'' crossings. A crossing is positive or negative depending on the orientation of the curves and which curve appears to be on top of the other. For closed curves the linking number is always an integer and is a topological invariant. If either of the curves is open, then the same approach can be used, but one must average over all projections of the curves rather than use a single projection. The linking number is not a topological invariant if either curve is open, and it can take any real value.

An alternative way to compute the linking number is the Gauss linking integral. For two (open or closed) curves $\gamma_1(s)$ and $\gamma_2(s')$, the linking number can be computed according to
\begin{equation}
    L[\gamma_1,\gamma_2] = \frac{1}{4\pi} \int_0^1 ds \int_0^1 ds' \, \frac{\det(\dot{\gamma}_1(s),\dot{\gamma}_2(s'),\gamma_1(s)-\gamma_2(s'))}{\left| \gamma_1(s) - \gamma_2(s') \right|^3}
\end{equation}
Notably the linking number can be positive or negative depending on the chosen orientation of the curves $\gamma_1$ and $\gamma_2$. All linear polymers considered here are head-tail symmetric, so there is no physical meaning to the sign of the linking number and we exclusively consider its absolute value $|L|$. The linking number has been generalized to periodic systems, in which it is known as the periodic linking number $L_P$ \cite{Panagiotou2015}.

 We first consider a model configuration between a ring and linear polymer in Figure \ref{fig:modelconfig} where the ring forms a perfect circle in a plane and the linear polymer originates inside the circle and extends in the direction orthogonal to the plane containing the circle. If the linear polymer were to extend infinitely away from the ring then $|L|=0.5$.

\begin{figure*}
    \centering
    \includegraphics{Figures/modelconfig.pdf}
    \caption{A model configuration between a ring polymer and a semi-infinite linear polymer viewed from three different projections: a) parallel to the linear and orthogonal to the plane containing the ring, b) parallel to the plane containing the ring, and c) from a perspective parallel to neither the plane containing the ring nor the direction of the linear polymer. The arrow indicates the direction in which the semi-infinite polymer continues.}
    \label{fig:modelconfig}
\end{figure*}

If the semi-infinite linear were to pass through the ring and originate from a point above the ring in Figure \ref{fig:modelconfig}b, then $|L|>0.5$. If the semi-infinite linear were to not pass through the ring and originate from a point below the ring in Figure \ref{fig:modelconfig}b then $|L|<0.5$. As such we use 0.5 as a cutoff value to determine if a ring-linear pair forms a thread.


Real linear and ring polymers have much more complicated configurations than those considered in Figure \ref{fig:modelconfig}, however simulations confirm that the thread definition cutoff of 0.5 is reasonable. Figure \ref{fig:GLIhist} shows the distribution of absolute periodic linking number $|L_P|$ values observed in a ring-linear blend with $N_R=N_L=200$ and $\phi_R=0.3$. The y-axis is the number of ring-linear pairs with a given $|L_P|$ value, normalized by the number of ring polymers in the blend.

\begin{figure}[h]
\includegraphics[scale=0.5]{Figures/GLIhist.pdf}
\caption{\label{fig:GLIhist} Distribution of absolute periodic linking number $|L_P|$ values in a ring-linear blend with composition $\phi_R=0.3$ and polymer lengths $N_R=N_L=200$. The y-axis is normalized by the number of rings in the blend.}
\end{figure}

The distribution is maximal near zero as most pairs of rings and linears are far apart in space and are not threaded. The distribution is also peaked around integer values and has local minima near 0.5, 1.5, etc. There are relatively few instances of pairs with $|L_P|\approx0.5$, making it a reasonable choice as a cutoff for threading.

Figure \ref{fig:GLIvtime} shows $|L_P|$ versus time for three selected ring-linear pairs from the same simulation as in Figure \ref{fig:GLIhist}. Time on the x-axis has been rescaled by the diffusion time of a linear chain.
For each pair of chains, $|L_P|$ tends to fluctuate around an integer value, but also undergoes rapid transitions between these states. Pair 1 (blue) is threaded at the start of the simulation, whereas the other pairs are not. At time $t/\tau_{d,L}=0.5$, pair 3 (green) becomes threaded and shortly after pair 2 (orange) also becomes threaded. At time $t/\tau_{d,L}=1$, pair 1 appears to become unthreaded, then rethreads a short time later. The absolute periodic linking number then jumps up to a value greater than 2 for a short period of time before returning to a value near unity, then finally jumps to a value near zero where it remains until the end of the simulation. These rapid transitions between states can also be observed in the other pairs.

\begin{figure}[h]
\includegraphics[scale=0.5]{Figures/GLIselect.pdf}
\caption{\label{fig:GLIvtime} Absolute periodic linking number $|L_P|$ for three selected ring-linear pairs versus time. Time has been rescaled by the diffusion time of a linear chain. The blend has composition $\phi_R=0.3$ and the polymers are length $N_R=N_L=200$. }
\end{figure}

Figures \ref{fig:GLIhist} and \ref{fig:GLIvtime} both show that $L_P$ can take values greater than unity and that ring-linear pairs can have configurations with $|L_P|\approx 2$, which indicate that one chain wraps around the other multiple times. These multiple threadings, though far less common that single threadings, may still be relevant to the dynamics of the system. We categorize threads according to the rounded absolute periodic linking number
\begin{equation}
    L_R = \left\lfloor |L_P| \right\rceil
\end{equation}
where $\lfloor X \rceil$ indicates to round $X$ to the nearest integer. Notably if a ring-linear pair is a thread based on the periodic linking number ($|L_P|>0.5$), then it will also be a thread based on the rounded absolute periodic linking number $L_R$. For counting and categorizing threads as single, double, triple, etc., $L_R$ proves convenient. Any ring-linear pair with $L_R>1$ we refer to as multiply threaded.

\begin{figure}[h]
    \centering
    \includegraphics{Figures/HigherOrderThreads_equalN.pdf}
    \caption{Average number of threads per ring with a given value of $L_R$ in a ring linear blend with $N_R=N_L$ and $\phi_R=0.3$. The thread multiplicity $L_R$ is indicated with the integer by each data series. The black line indicates a linear slope $\sim N$.}
    \label{fig:multithread}
\end{figure}

Figure \ref{fig:multithread} shows the average number of threads per ring according to the thread multiplicity $L_R$ in a blend with $\phi_R=0.3$ and $N_R=N_L$. Threads with $L_R$ up to seven were observed. The number of single threads $L_R=1$ increases linearly with $N$ and is the most common type of thread. The number of multiple threads per ring decreases exponentially with increasing $L_R$, i.e. $\sim e^{-\gamma L_R}$, where $\gamma$ is a fit parameter. Nonetheless, the number of multiple threads ($L_R>1$) increases faster than linearly with $N_R$, though does not seem to follow a power scaling.

\newpage

\section{Re-entrant threads}
We define a re-entrant thread as a ring-linear pair for which $|L_P|<0.5$, but with $|L_P|>0.5$ if only a subchain of the linear chain is considered. Examples of re-entrant threads are shown in Figure \ref{fig:reentrant}. Figure \ref{fig:reentrant}(a) shows a ring-linear pair with $|L_P|<0.5$. If one considers only the subchain of the linear polymer from the leftmost end to the middle segment, then $|L_P|>0.5$. There are many such conformations observed in the MD simulations, however we do not expect such conformations to be relevant to the dynamics as the linear subchain threads the ring very shallowly. To remove the re-entrant thread only a few segments in the middle of the red chain must move down. In contrast, Figure \ref{fig:reentrant}(b) shows a re-entrant thread (red and black curves) that is deeper than in part (a). Furthermore the red linear chain is entangled with the blue linear chain. Together the three chains form a topological entanglement that can not be easily removed via the motion of a few segments of any polymer.

It is thus ambiguous from the periodic linking number alone whether a re-entrant thread constitutes a relevant topological constraint. However, most re-entrant threads are shallow as in Figure \ref{fig:reentrant}(a) and there is fewer than one re-entrant thread per ring that has length greater than the entanglement length. We thus neglect \emph{all} re-entrant threads in our analysis.

\begin{figure}
\includegraphics[scale=0.5]{Figures/Reentrant.pdf}
\caption{\label{fig:reentrant} Examples of re-entrant threads. (a) A shallow re-entrant thread that can be removed by the motion of a few segments in the middle of the red linear chain. (b) A deep re-entrant thread that forms a three-chain topological entanglement.}
\end{figure}

\newpage

\section{Dethreading correlation function}
We define a correlation function to measure dethreading. Consider a melt of $M_R$ ring polymers and $M_L$ linear polymers where ring polymer $i$ is defined by its space curve $\gamma_i(s,t)$ and linear polymer $j$ is defined by its space curve $\eta_j(s',t)$ with $s,s'\in [0,1]$ parameterizing the chain contour position and $t$ representing time. One way to define the correlation function is
\begin{equation}
	\tilde{C}(t) = \frac{\left\langle \sum\limits_{i=1}^{M_R} 
 \sum\limits_{j=1}^{M_L} 
 \Theta(\left|L_P[\gamma_i(s,t),\eta_j(s',t)]\right| -0.5 ) 
 \Theta(\left|L_P[\gamma_i(s,0),\eta_j(s',0)]\right| -0.5 )
 \right\rangle}
 {\left\langle 
 \sum\limits_{i=1}^{M_R} 
 \sum\limits_{j=1}^{M_L} 
 \Theta(\left|L_P[\gamma_i(s,0),\eta_j(s',0)]\right| -0.5 )
 \right\rangle}
\end{equation}
where $\Theta$ is the heaviside function. The denominator represents the total number of ring-pairs that are threaded ($|L_P|>0.5$) at time $t=0$. The numerator represents the total number of ring-linear pairs that are threaded at both time $t$ and at the initial time.
The angle brackets indicate an average over time origins. 
The correlation function is constrained to values in $[0,1]$ and will typically decrease from unity to zero monotonically. Notably, it is possible for ring-linear pairs to become unthreaded at some time in $(0,t)$, then become rethreaded by time $t$. Thus, the correlation function can be thought of as a measure of the terminal relaxation time for a thread. The time at which the correlation function achieves a value of $0.5$ corresponds to the median final escape time. One deficiency with this definition is that there are pairs of chains that are multiply threaded (such as pair 2 at the final time in Fig. \ref{fig:GLIvtime}). This correlation function only responds when the pair becomes completely unthreaded and does not capture the release of one of the multiple threads. To account for this we modify the correlation function to
\begin{equation}
	C(t) = \frac{\left\langle \sum\limits_{i=0}^{M_R} 
 \sum\limits_{j=0}^{M_L} 
 L_R[\gamma_i(s,t),\eta_j(s',t)] 
 \Theta(\left|L_P[\gamma_i(s,0),\eta_j(s',0)]\right| -0.5 )
 \right\rangle}
 {\left\langle 
 \sum\limits_{i=0}^{M_R} 
 \sum\limits_{j=0}^{M_L} 
 L_R[\gamma_i(s,0),\eta_j(s',0)]
 \right\rangle}
\end{equation}
By using $L_R$ rather than $\Theta(|L_P|-0.5)$ we account for multiple threads.
With this definition, the correlation function will partially decay when a multiple thread releases one of its threads.

The correlation functions presented thus far describe the final relaxation of ring-linear threads. We present here an alternative correlation function that describes the first relaxation time of ring-linear threads, which is related to the continuous association correlation function. The correlation function is defined as
\begin{equation}
	\tilde{S}(t) = \frac{ \left\langle
 \sum\limits_{i=0}^{M_R} \sum\limits_{j=0}^{M_L} 
 \prod\limits_{t'\in T}
\Theta(\left|L_P[\gamma_i(s,t'),\eta_j(s',t')]\right| -0.5 )
 \right\rangle }
  {\left\langle 
 \sum\limits_{i=0}^{M_R} 
 \sum\limits_{j=0}^{M_L} 
 \Theta(\left|L_P[\gamma_i(s,0),\eta_j(s',0)]\right| -0.5 )
 \right\rangle}
\end{equation}
where $T=\{0,\delta t,2\delta t,\ldots,t\}$ is the set of all time points between $0$ and $t$ measured in intervals of $\delta t \leq t$. This continuous association correlation function $\tilde{S}(t)$ tracks the set of ring-linear pairs that remain threaded for all times between $0$ and $t$. This is in contrast to $C(t)$ which only requires that the ring-linear pair is threaded at the start and end point of the time interval. The correlation $\tilde{S}(t)$ is monotonically non-increasing while $C(t)$ is not. Additionally, $\tilde{S}(t) \leq \tilde{C}(t)$ for all $t$. The exact value of $\tilde{S}(t)$ depends on the choice of $\delta t$. Typically $\delta t$ should be converged to a sufficiently small value so that $\tilde{S}(t)$ no longer depends on $\delta t$. This can be achieved by choosing $\delta t$ smaller than the shortest time for a ring-linear pair to unthread and rethread. In practice, this time can be very small, especially for small polymers, such that $\delta t$ would be orders of magnitude smaller than other time scales relevant to chain motion and material properties. As such we typically do not consider $\tilde{S}(t)$, but rather focus on $C(t)$ which indicates the final dethreading of a ring-linear pair. The first dethreading time captured by $\tilde{S}(t)$ does not account for chains that remain near each other and rethread due to other topological constraints. These dethreading events do not correspond to an actual stress relaxation event in the system. In contrast, dethreading events captured by $C(t)$ likely indicate a true reconfiguration of the chains that may allow for stress relaxation.

Furthermore, $\tilde{S}(t)$ is susceptible to errors from re-entrant thread formation. A thread may become a re-entrant thread that still forms a topological constraint, however $\tilde{S}(t)$ will view this as an unthreading. As previously discussed, these re-entrant threads are not common, so at a given time point it will have a small effect on $C(t)$. The continuous association correlation function $S(t)$ depends on the entire time window $[0,t]$, so the likelihood of an error increases as $t$ increases. As such, we view $C(t)$ as a more reliable measure of the topological entanglements in the blend.
\newpage
\section{Mean-squared displacement}
Examples of  mean squared displacement of the center of mass of the chains $g_3(t)$ and  of the beads $g_1(t)$ is shown in Fig. \ref{fig:ringMSD} for rings of length $N_R=400$ in a pure ring melt and for rings in a blend with $\phi_R=0.3$  for varying linear chain length $100 \le N_L \le 600$.  Figure \ref{fig:linearMSD} shows the same two quantities for a linear chain of length $N_L=400$ in a pure linear melt and in a blend with rings of length $N_R=400$ for $\phi_R=0.3$ and $0.5$.
Here $g_1(t)$ is averaged over the inner 10 beads of the linear chain.

\begin{figure}
    \centering
    \includegraphics[scale=0.4]{Figures/MSD_rings_R400-L100-600.pdf}
    \caption{Mean-squared displacement of ring polymers of length $N_R=400$ in pure melts (circles) or blends with $\phi_R=0.3$ (all other symbols). Mean squared displacement of the center of  mass $g_3(t)$ is shown in blue and motion of a monomer $g_1(t)$ in black.}
    \label{fig:ringMSD}
\end{figure}

\begin{figure}
    \centering
    \includegraphics[scale=0.4]{Figures/MSD-400-LINEAR IN BLEND.pdf}
    \caption{Mean-squared displacement of linear polymers in pure melts (black squares) or a blend with $\phi_R=0.3$ (red circles) and $0.50$ (blue triangles). Mean squared placement of the center of  mass $g_3(t)$ (open) and motion of a monomer $g_1(t)$ (filled). }
    \label{fig:linearMSD}
\end{figure}

\section{Crossover functions}
We fit the data for $\tau_{d,R}$ vs $N_L$ with $N_R$ fixed to a crossover function of the form
\begin{equation}
    \tau_{d,R} = \tau_{d,R0}\left(1+\left(\frac{N_L}{N_{L,c}}\right)^\frac{3.4}{\alpha}\right)^\alpha
    \label{eq:tau_fit}
\end{equation}
here $N_{L,c}$ represents a crossover linear chain length. The parameters for fits to the datasets with $\phi_R=0.3$ and $N_R=200$ or $N_R=400$ are listed in Table \ref{tab:tauD_fit}.

\begin{table}[h]
    \centering
        \caption{Fit parameters for crossover function for $\tau_{d,R}$ vs $N_L$ for blends with $\phi_R=0.3$ and various fixed $N_R$.}
    \begin{tabular}{c | c c c}
        \hline
         $N_R$ & $\tau_{d,R0}/\tau$ & $N_{L,c}$ & $\alpha$  \\
        \hline
        200 & $11661\pm\ 32$ & $93.6\pm1.0$ & $3.49\pm0.06$ \\
        400 & $30989\pm224$ & $68.4\pm1.8$ & $3.66\pm0.13$ \\
        \hline
    \end{tabular}
    \label{tab:tauD_fit}
\end{table}

We also fit the data for the ring radius of gyration $R_g$ versus $N_R$ to a crossover function
\begin{equation}
    R_g^2 = \frac{2R_{gc}^2}{\left(\left(\frac{N_R}{N_{Rc}}\right)^{-\beta} + \left(\frac{N_R}{N_{Rc}}\right)^{\frac{-2\beta}{3}}\right)^{\frac{1}{\beta}}} \label{eq:Rgfit}
\end{equation}
The parameters for the fits to equation \ref{eq:Rgfit} are given in Table \ref{tab:Rg_fit}.

\begin{table}[h]
    \centering
    \begin{tabular}{c | c c c}
        \hline
         $\phi_R$ & $R^2_{g0}/\sigma^2$ & $N_{Rc}$ & $\beta$  \\
        \hline
        0.3 & $44.8\pm0.3$ & $445.2\pm\ 3.2$ & $6.52\pm0.22$ \\
        0.5 & $34.2\pm0.3$ & $328.6\pm\ 4.1$ & $3.04\pm0.12$ \\
        1.0 & $13.3\pm0.8$ & $122.0\pm11.2$ & $2.67\pm0.14$ \\
        \hline
    \end{tabular}
    \caption{Fit parameters for crossover function for ring $R_g$ vs $N_R$ for pure rings and ring-linear blends with various $\phi_R$. Blends have $N_L=200$.}
    \label{tab:Rg_fit}
\end{table}

\newpage

\bibliography{references}